# Timeline: A Dynamic Hierarchical Dirichlet Process Model for Recovering Birth/Death and Evolution of Topics in Text Stream

**Amr Ahmed**
amahmed@cs.cmu.edu
School of Computer Science
Carnegie Mellon University

**Eric P. Xing**
epxing@cs.cmu.edu
School of Computer Science
Carnegie Mellon University

## Abstract

Topic models have proven to be a useful tool for discovering latent structures in document collections. However, most document collections often come as temporal streams and thus several aspects of the latent structure such as the number of topics, the topics' distribution and popularity are time-evolving. Several models exist that model the evolution of some but not all of the above aspects. In this paper we introduce *infinite* dynamic topic models, iDTM, that can accommodate the evolution of all the aforementioned aspects. Our model assumes that documents are organized into epochs, where the documents within each epoch are exchangeable but the order between the documents is maintained across epochs. iDTM allows for unbounded number of topics: topics can die or be born at any epoch, and the representation of each topic can evolve according to a Markovian dynamics. We use iDTM to analyze the birth and evolution of topics in the NIPS community and evaluated the efficacy of our model on both simulated and real datasets with favorable outcome.

## 1 Introduction

With the dramatic increase of digital document collections such as online journal articles, the Arxiv, conference proceedings, blogs, to name a few, there is a great demand for developing automatic text analysis models for analyzing these collections and organizing its content. Statistical admixture topic models (Blei et al., 2003) were proven to be a very useful tool to attain that goal and have recently gained much popularity in managing large collection of documents. Via an admixture model, one can project each document into a low dimensional space where their latent semantic (such as topical aspects) can be captured. This low dimensional representation can then be used for tasks like measuring document-document similarity or merely as a visualization tool that gives a bird's eye view of the collection and guides its exploration in a structured fashion.

An admixture topic model posits that each document is sampled from a fixed-dimensional mixture model according to a document's specific mixing vector over the *topics*. The variabilities in the topic mixing vectors of the documents are usually modeled as a Dirichlet distribution (Blei et al., 2003), although other alternatives have been explored in the literature (Blei and Lafferty, 2007, Li and McCallum, 2006). The components of this Dirichlet distribution encode the popularity of each of the topics in the collection. However, document collections often come as temporal streams where documents can be organized into epochs; examples of an epoch include: documents in an issue of a scientific journal or the proceeding of a conference in a given year. Documents inside each epoch are assumed to be exchangeable while the order between documents is maintained across epochs. With this organization, several aspects of the aforementioned static topic models are likely to change over time, specifically: topic *popularity*, topic *word distribution* and the *number* of topics.

Several models exist that could accommodate the evolution of some but not all of the aforementioned aspects. Blei and Lafferty (2006) proposed a dynamic topic model in which the topic's word distribution and popularity are linked across epochs using state space models, however, the number of topics are kept fixed. Wang and McCallum (2006) presented the topics over time model that captures topic popularity over time via a beta distribution, however, topic distributions over words and the number of topics were fixed over time, although the authors discussed a non-parametric extension over the number of topics. Moreover, the shapes of permitted topic trends in the TOT model are restricted to those attained by the beta distribution. On the other hand, several models were proposed that could *potentially* evolve all the aforementioned aspect albeit in a simple clustering settings, i.e. each document is assumed to be sampled from a single topic (Ahmed and Xing, 2008, Caron et al., 2007, Srebro and Roweis, 2005). As we will show in this paper, accommodating the evolution of the aforementioned aspects in a full-fledged admixture setting is non-trivial and introduces its own hurdles. Moreover, it is widely accepted (Blei et al., 2003) that admixture models are superior compared to simple clustering models for modeling text documents, especially for long documents such as research papers.

In this paper we introduce iDTM: infinite dynamic topic models which can accommodate the evolution of the aforementioned aspects. iDTM allows for unbounded number of topics: topics can be born and die at any epoch, the topics' word distributions evolve according to a first-order state space model, and the topics' popularity evolve using the rich-gets richer scheme via a $\Delta$-order process. iDTM is built on top of the recurrent Chinese restaurant franchise (RCRF) process which introduces dependencies between the atom locations (topics) and weights (popularity) of each epoch. The RCRF process is built on top of the RCRP process introduce in (Ahmed and Xing, 2008).

To summarize, the contributions of this paper are:

- A principled formulation of a dynamic topic model that evolves: topic trend, topic distribution, and number of topics over time.
- An efficient sampling algorithm that relies on dynamic maintenance of cached sufficient statistics to speed up the sampler.
- An empirical evaluation and illustration of the proposed model over simulated data and over the NIPS proceedings.
- A study of the sensitivity of the model to the setting of its hyperparameters.

## 2 Settings and Background

In this section, we lay the foundation for the rest of this paper by first detailing our settings and then reviewing some necessary background to make the paper self-contained. We are interested in modeling an ordered set of documents $\boldsymbol{w} = (\boldsymbol{w_1}, \cdots, \boldsymbol{w_T})$, where $T$ denotes the number of epochs and $\boldsymbol{w_t}$ denotes the documents at epoch $t$. Furthermore, $\boldsymbol{w_t} = (\boldsymbol{w_{td}})_{d=1}^{D_t}$, where $D_t$ is the number of documents at epoch $t$. Moreover, each document comprises a set of $n_{td}$ words, $\boldsymbol{w_{td}} = (w_{tdi})_{i=1}^{N_{td}}$, where each word $w_{tdi} \in \{1, \cdots, W\}$. Our goal is to discover *potentially* an unbounded number of topics $(\boldsymbol{\phi}_k)_{k=1}^\infty$ where each topic $\boldsymbol{\phi}_k = (\phi_{k,t_{k_1}}, \cdots, \phi_{k,t_{k_2}})$ spans a set of epoches where $1 \leq t_{k_1} \leq t_{k_2} \leq T$, and $\phi_{k,t}$ is the topic's word distribution at epoch $t$.

### 2.1 Temporal Dirichlet Process Mixture Model

The Dirichlet process (DP) is a distribution over distributions (Ferguson, 1973). A DP denoted by $DP(G_0, \alpha)$ is parameterized by a base measure $G_0$ and a concentration parameter $\alpha$. We write $G \sim DP(G_0, \alpha)$ for a draw of a distribution $G$ from the Dirichlet process. $G$ itself is a distribution over a given parameter space $\theta$, therefore we can draw parameters $\theta_{1:N}$ from $G$. Integrating out $G$, the parameters $\boldsymbol{\theta}$ follow a Polya urn distribution (Blackwell and MacQueen, 1973), also knows as a Chinese restaurant process (CRP), in which the previously drawn values of $\theta$ have strictly positive probability of being redrawn again, thus making the underlying probability measure $G$ discrete with probability one. More formally,

$$\theta_i | \theta_{1:i-1}, G_0, \alpha \sim \sum_k \frac{m_k}{i-1+\alpha} \delta(\phi_k) + \frac{\alpha}{i-1+\alpha} G_0. \quad (1)$$

where, $\phi_{1:k}$ denotes the distinct values among the parameters $\boldsymbol{\theta}$, and $m_k$ is the number of parameters $\theta$ having value $\phi_k$. By using the $DP$ at the top of a hierarchical model, one obtains the Dirichlet process mixture model, DPM (Antoniak, 1974).

Several approached have been proposed to introduce temporal dependencies in DPs (Ahmed and Xing, 2008, Caron et al., 2007, Ghahramani and Lafferty, 2005, Griffn and Steel, 2006, Rao and Teh , 2009), to name a few. Here we focus on the temporal DPM introduced in (Ahmed and Xing, 2008). The temporal Dirichlet process mixture model (TDPM) is a framework for modeling complex longitudinal data, in which the number of mixture components at each time point is unbounded; the components themselves can retain, die out or emerge over time; and the actual parameterization of each component can also evolve over time in a Markovian fashion. In TDPM, the random measure $G$ is time-varying, and the process stipulates that:

$$G_t | \phi_{1:k}, G_0, \alpha \sim \quad (2)$$
$$DP\left(\alpha + \sum_k m'_{kt}, \sum_k \frac{m'_{k,t}}{\sum_l m'_{lt} + \alpha} \delta(\phi_k) + \frac{\alpha}{\sum_l m'_{lt} + \alpha} G_0\right)$$

where $\{\phi_{1:k}\}$ are the mixture components available in the previous $\Delta$ epochs, in other words, $\{\phi_{1:k}\}$ is the collection of unique values of the parameters $\boldsymbol{\theta}_{t:t-\Delta}$, where $\theta_{tn}$ is the parameter associated with data point $x_{tn}$. If we let $m_{kt}$ denotes the number of parameters in epoch $t$ associated with component $k$, then $m'_{kt}$, the prior weight of component $k$ at epoch $t$ is defined as:

$$m'_{kt} = \sum_{\delta=1}^{\Delta} \exp^{\frac{-\delta}{\lambda}} m_{k,t-\delta} \quad (3)$$

,where $\Delta, \lambda$ define the width and decay factor of the time-decaying kernel. This defines a $\Delta-$order process where the strength of dependencies between epoch-specific DPs are controlled with $\Delta, \lambda$. Ahmed and Xing (2008) showed that these two hyper-parameters allow the TDPM to degenerate to either a set of independent DPMs at each epoch when $\Delta=0$, and to a global DPM, i.e ignoring time, when $\Delta = T$ and $\lambda = \infty$. In between, the values of these two parameters affect the expected life span of a given component. The larger the value of $\Delta$ and $\lambda$, the longer the expected life span of the topic, and vice versa. Finally, the life-span of topics followed a power law distribution (Ahmed and Xing, 2008).

In addition to changing the weight associated with each component, the parameterization $\phi_k$ of each component changes over time in a markovian fashion, i.e.: $\phi_{kt}|\phi_{k,t-1} \sim P(.|\phi_{k,t-1})$. Integrating out the random measures $G_{1:T}$, the parameters $\boldsymbol{\theta}_{1:t}$ follow a polya-urn distribution with time-decay, or as termed in Ahmed and Xing (2008), the recurrent Chinese restaurant process (RCRP). More formally:

$$\theta_{ti}|\boldsymbol{\theta_{t-1:t-\Delta}}, \theta_{t,1:i-1}, G_0, \alpha \propto \sum_k \left(m'_{kt} + m_{kt}\right)\delta(\phi_{kt})$$
$$+ \alpha G_0 \quad (4)$$

The RCRP allows each document **w** to be generated from a single component (topic), thus making it suboptimal in modeling multi-topic documents. In the next subsection, we review the HDP process which allows each document to be generated from multiple topics.

### 2.2 Hierarchical Dirichlet Processes

Instead of modeling each document $\mathbf{w}_d$ as a single data point, we could model each document as a DP. In this setting, each word $w_{dn}$ is a data point and thus will be associated with a topic sampled from the random measure $G_d$, where $G_d \sim DP(\alpha, G_0)$. The random measure $G_d$ thus represents the document-specific mixing vector over a potentially infinite number of topics. To share the same set of topics across documents, Teh et al. (2006) introduced the Hierarchical Dirichlet Process (HDP). In HDP, the document-specific random measures are tied together by modeling the base measure $G_0$ itself as a random measure sampled from a $DP(\gamma, H)$. The discreteness of the based measure $G_0$ ensures sharing of the topics between all the groups.

Integrating out all random measures, we obtain the equivalent Chinese restaurant franchise processes(CRFP) (Teh et al., 2006). In our document modeling setting, the generative story behind this process proceeds as follows. Each document is refereed to as a restaurant where words inside the document are referred to as customers. The set of documents shares a global menu of topics. The words in each document are divided into groups, each of which shares a table. Each table is associated with a topic (dish in the metaphor), and words sitting on each table are associated with the table's topic. To associate a topic with word $w_{di}$ we proceed as follows. The word can sit on table $b_{db}$ that has $n_{db}$ words with probability $\frac{n_{db}}{i-1+\alpha}$, and shares the topic, $\psi_{db}$ on this table, or picks a new table, $b^{\text{new}}$ with probability $\frac{\alpha}{i-1+\alpha}$ and orders a new topic, $\psi_{db^{\text{new}}}$ sampled from the global menu. A topic $\phi_k$ that is used in $m_k$ tables across all documents is ordered from the global menu with probability $\frac{m_k}{\sum_{l=1}^{K} m_l + \gamma}$, or a new topic, $k^{\text{new}}$ not used in any document, is ordered with probability $\frac{\gamma}{\sum_{l=1}^{K} m_l + \gamma}$, $\phi_{k^{\text{new}}} \sim H$. If we let $\theta_{di}$ denotes the distribution of the topic associated with word $w_{di}$, then putting everything together we have:

$$\theta_{di}|\theta_{d,1:i-1}, \alpha, \boldsymbol{\psi} \sim \sum_{b=1}^{b=B_d} \frac{n_{db}}{i-1+\alpha}\delta_{\psi_{db}} + \frac{\alpha}{i-1+\alpha}\delta_{\psi_{db^{\text{new}}}} \quad (5)$$

$$\psi_{db^{\text{new}}}|\boldsymbol{\psi}, \gamma \sim \sum_{k=1}^{K} \frac{m_k}{\sum_{l=1}^{K} m_l + \gamma}\delta_{\phi_k} + \frac{\gamma}{\sum_{l=1}^{K} m_l + \gamma} H \quad (6)$$

where $B_d$ is the number of tables in document $d$.

While the CRFP allows each document to be generated from multiple topics, and allows for the number of topics to be unbounded, it still can not evolve the topics' trends and word distributions.

## 3 Infinite Dynamic Topic Models

Now we proceed to introducing our model, iDTM which allows for infinite number of topics with variable durations. The documents in epoch $t$ are modeled using an epoch specific HDP with high-level base measure denoted as $G_0^t$. These epoch-specific base measures $\{G_0^t\}$ are tied together using the TDPM process of (Ahmed and Xing, 2008). Integrating all random measure, we get the recurrent Chinese restaurant franchise process (RCRF).

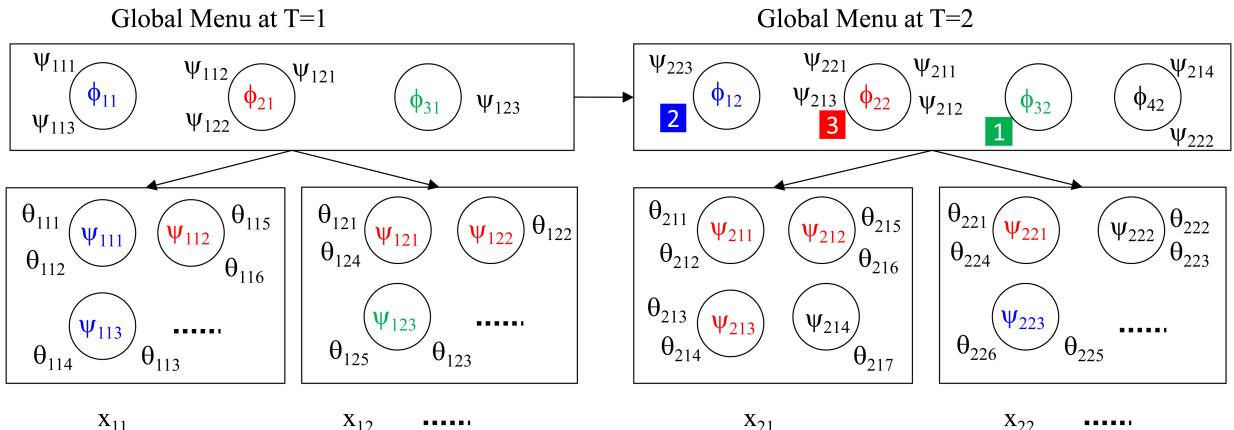

Figure 1: The recurrent Chinese restaurant franchise (RCRF) precoces. The figure shows a first-order process with no decay to avoid cluttering the display, however see the text for the description of a general $\Delta$-order process.

Figure 1 depicts a RCRF process of order one for clarity, however, in this section we give a description of a general process of order $\Delta$. In the RCRF, the document in each epoch is modeled using a CRFP, and then the global menus of each epoch are tied over time as depicted in Figure 1.

As in the RCRP, the popularity of a topic at epoch $t$ depends both on its usage at this epoch, $m_{kt}$ as well as it historic usage at the proceedings $\Delta$ epochs, $m'_{kt}$, where $m'_{kt}$ is given in (3). This means that a topic is considered *dead* only when it is unused for a consecutive $\Delta$ epochs. For simplicity, we let $m'_{kt} = 0$ for newly-born topics at epoch $t$, and $m_{kt} = 0$ for topics available to be used(i.e having $m'_{kt} > 0$) but not yet used in any document at epoch $t$.

The generative process at the first epoch proceeds exactly as in the CRF process. At epoch $t$, to associate a topic with word $w_{tdi}$ we proceed as follows. Word $w_{tdi}$ can sit on table $b$ that has $n_{tdb}$ customers and has topic $\psi_{tdb}$ with probability $\frac{n_{tdb}}{i-1+\alpha}$. Alternatively, $w_{tdi}$ can choose to start a new table, $b_{td}^{new}$ with probability $\frac{\alpha}{i-1+\alpha}$ and choose a new topic. It can choose an already existing topic from the menu at epoch $t$ with probability $\frac{m_{kt}+m'_{kt}}{\sum_{l=1}^{K_t} m_{lt}+m'_{lt}+\gamma}$, $K_t$ is the number of topics at epoch $t$. Furthermore, if this topic is inherited but not yet used by any previous word (i.e $m_{kt} = 0$), then $w_{tdi}$ modifies the distribution of this topic: $\phi_{kt} \sim P(.|\phi_{k,t-1})$. Finally, $w_{tdi}$ can choose a brand new topic $\phi_{K_t^{new}} \sim H$, with probability $\frac{\gamma}{\sum_{l=1}^{K_t} m_{lt}+m'_{lt}+\gamma}$ and increment $K_t$. Putting everything together, we have:

$$\theta_{tdi}|\theta_{td,1:i-1},\alpha,\psi_{t-\Delta:t} \sim \sum_{b=1}^{b=B_{td}} \frac{n_{tdb}}{i-1+\alpha}\delta_{\psi_{tdb}} + \frac{\alpha}{i-1+\alpha}\delta_{\psi_{tdb^{new}}} \quad (7)$$

$$\psi_{tdb^{new}}|\psi,\gamma \sim \sum_{k:m_{kt}>0} \frac{m_{kt}+m'_{kt}}{\sum_{l=1}^{K_t} m_{lt}+m'_{lt}+\gamma}\delta_{\phi_{kt}}$$
$$+ \sum_{k:m_{kt}=0} \frac{m_{kt}+m'_{kt}}{\sum_{l=1}^{K_t} m_{lt}+m'_{lt}+\gamma}P(.|\phi_{k,t-1})$$
$$+ \frac{\gamma}{\sum_{l=1}^{K_t} m_{lt}+m'_{lt}+\gamma} H \quad (8)$$

If we use the RCRF process as a prior over word assignment to topics in a mixed-membership model, we get the infinite dynamic topic model (iDTM). In iDTM, each word is assigned to a topic as in the RCRF process, and then the word is generated from this topic's distribution. The base measure $H$ is modeled as $H = N(0, \sigma I)$, and the word distribution of the topic $k$ at epoch $t$, $\phi_{kt}$ evolves using a random walk kernel as in (Blei and Lafferty, 2006): $\phi_{k,t}|\phi_{k,t-1} \sim N(\phi_{k,t-1}, \rho I)$. To generate word $w_{tdi}$ from its associated topic, say $\phi_{kt}$, we first map the natural parameters of this topic $\phi_{k,t}$ to the simplex via the logistic transformation $L$, and then generate the word, i.e.: $w_{tdi}|\phi_{kt} \sim \mathcal{M}(L(\phi_{kt}))$, where $L(\phi_{kt}) = \frac{exp^{\phi_{kt}}}{\sum_{w=1}^{W} exp^{\phi_{k,t,w}}}$. This choice introduces non-conjugacy between the base measure and the likelihood function which we have to deal with in Section 4.

## 4 A Gibbs Sampling Algorithm

In this section, we give a Gibbs sampling algorithm for posterior inference in the iDTM. We construct a Markov chain over $(\mathbf{k}, \mathbf{b}, \boldsymbol{\phi})$, where $k_{tdb}, b_{tdi}, \phi_{kt}$ are as given in Section 3: the index of the topic on table $b$ in document $td$, the table index assigned to word $w_{tdi}$, and the parameterization of topic $k$ at time epoch $t$, respectively. We use the following notations. Adding a superscript $-i$ to a variable, indicate the same quantity it is added to without the contribution of object $i$. For example $n_{tdb}^{-tdi}$ is the number of customers sitting on table $b$ in document $d$ in epoch $t$ without the contribution of word $w_{tdi}$, and $\boldsymbol{k}_t^{-tdb}$ is $\boldsymbol{k}_t \backslash k_{tdb}$. We alternate sampling each variable conditioned on its Markov blanket as follows:

**Sampling a topic $k_{tdb}$ for table $tdb$:** The conditional distribution for $k_{tdb}$ is given by:

$$P(k_{tdb} = k | \boldsymbol{k}_{t-\Delta:t+\Delta}^{-tdb}, \boldsymbol{b}_{td}, \boldsymbol{\phi}, \boldsymbol{w}_t) \propto \qquad (9)$$
$$P(k_{tdb} = k | \boldsymbol{k}_{t-\Delta:t}^{-tdb}, \boldsymbol{\phi}, \boldsymbol{v}_{tdb}) \prod_{\delta=1}^{\Delta} P(\boldsymbol{k}_{t+\delta} | \boldsymbol{k}_{t+\delta-\Delta:t+\delta-1}^{-tdb \to k})$$

where $\boldsymbol{v}_{tdb}$ is the frequency count vector (of length W) of the words sitting on table $tdb$, and the notation $(^{-tdb \to k})$ means the same as $\boldsymbol{k}_{t-\Delta:t+\Delta}^{-tdb}$ but in addition we set $k_{tdb} = k$. The second factor in (9) is the transition probability which measures the likelihood of the table assignments at future epochs if we choose to assign $k_{tdb} = k$. Now we focus on computing one of these probabilities in the second factor in (9) at epoch $t+\delta$. With reference to the construction in Section 3 and Eq (8), and considering that documents are exchangeable within each epoch, similar to Antoniak (1974), we have [1]:

$$P(\boldsymbol{k}_{t+\delta} | \boldsymbol{k}_{t+\delta-\Delta:t+\delta-1}^{-tdb \to k}) = \qquad (10)$$
$$\gamma^{K_{t+\delta}^{born}} \frac{\prod_{s \in K_{t+\delta}^{born}} [1]^{m_{s,t+\delta}} \prod_{s \notin K_{t+\delta}^{born}} [m_{s,t+\delta}^{',-tdb \to k}]^{m_{s,t+\delta}}}{\prod_{i=1}^{m_{\cdot,t+\delta}} (m_{\cdot,t+\delta}^{',-tdb \to k} + \gamma + i)}$$

where $K_{t+\delta}^{born}$ is the number of topics born at epoch $t+\delta$, $m_{\cdot,t+\delta}$ is the summation of $m_{k,t+\delta}$ over the first dimension (the topic), and $m'_{\cdot,t+\delta}$ is defined similarly. Finally, $[a]^c = a(a+1) \cdots (a+c-1)$.

Now we turn to the first factor in (9). Using (8), $P(k_{tdb} = k | \boldsymbol{k}_{t-\Delta:t}^{-tdb}, \boldsymbol{\phi}, \boldsymbol{v}_{tdb}) \propto$ :

$$\begin{cases} (m_{kt}^{-tdb} + m'_{kt}) f(\boldsymbol{v}_{tdb} | \phi_{kt}) \\ \qquad \text{k is being used}: m_{kt}^{-tdb} > 0 \\ m'_{kt} \int f(\boldsymbol{v}_{tdb} | \phi_{kt}) dP(\phi_{kt} | \phi_{k,t-1}) \\ \qquad \text{k is available but not used}: m'_{kt} > 0 \\ \gamma \int f(\boldsymbol{v}_{tdb} | \phi_{kt}) dH(\phi_{kt}) \\ \qquad \text{k is a new topic} \end{cases} \quad (11)$$

Unfortunately, due to the non-conjugacy neither the second nor the third case above can be computed analytically. In Ahmed and Xing (2008) a Laplace approximation was used to fit these integrals. This was possible since the integrals were evaluated over the whole document (a mixture model), however in our setting (mixed-membership model), we need to evaluate these integrals over small groups of words (like words on a given table). We found that the deterministic approximation overestimates the integrals and increases the rate of generating new topics, therefore we resort to algorithm 8 in Neal (1998). In this case, we replace both of these integrals with $Q$ fresh samples from their respective distributions, and equally divide the corresponding probability mass among these new samples. These samples are then treated as if they were already existing topics. We choose to use $Q = 1$ for the transition kernel since in our application, iDTM, this kernel usually has a small variance. Substituting this in equation (11), we get that $P(k_{tdb} = k | \boldsymbol{k}_{t-\Delta:t}^{-tdb}, \boldsymbol{\phi}, \boldsymbol{v}_{tdb}) \propto$:

$$\begin{cases} (m_{kt}^{-tdb} + m'_{kt}) f(\boldsymbol{v}_{tdb} | \phi_{kt}) \\ \qquad \text{k is used}: m_{kt}^{-tdb} > 0 \\ m'_{kt} f(\boldsymbol{v}_{tdb} | \phi_{kt}) \\ \qquad m'_{kt} > 0, m_{kt}^{-tdb} = 0, \phi_{kt} \sim P(.|\phi_{k,t-1}) \\ \frac{\gamma}{Q} f(\boldsymbol{v}_{tdb} | \phi_{kt}^q) \\ \qquad \text{k is a new topic}, \phi_{kt}^q \sim H, q = 1 \cdots Q \end{cases} \quad (12)$$

**Sampling a table $b_{tdi}$ for word $x_{tdi}$:** With reference to (7), the conditional distribution for $b_{tdi}$, $P(b_{tdi} = b | \boldsymbol{b}_{td}^{-tdi}, \boldsymbol{k}_{t-\Delta:t+\Delta}, \boldsymbol{\phi}, x_{tdi})$ is proportional to:

$$\begin{cases} n_{tdb}^{-tdi} f(x_{tdi} | \phi_{k_{tjb},t}) \\ \qquad \text{b is an existing table} \\ \alpha P(k_{tjb^{new}} = k | \boldsymbol{k}_{t-\Delta:t+\Delta}^{-tdi}, \boldsymbol{b}_{td}^{-tdi}, \boldsymbol{\phi}, x_{tdi},) \\ \qquad b^{new} \text{ is a new table}, k \in \boldsymbol{k}_t \end{cases} \quad (13)$$

Several points are in order to explain (13). There are two choices for word $w_{tdi}$: either to sit on an existing table, or to sit on a new table and choose a new topic. In the second case, we need to sample a topic for this new table which leads to the same equation as in (11). Moreover, if $w_{tdi}$ was already sitting on a table by itself, then we need to first remove the contribution of this table from the count vector $\boldsymbol{m}$. Finally, note that in the second line, $P$ is a proper distribution (i.e. it should be normalized) and thus the total probability mass for sitting on a new table is till $\alpha$ regardless of how many topics are available at epoch $t$.

**Sampling $\phi_k$:** $P(\phi_k | \boldsymbol{b}, \boldsymbol{k}, \boldsymbol{x}) = P(\phi_k | \boldsymbol{v}_k)$, where $\boldsymbol{v}_k = \{\boldsymbol{v}_{k,t}\}$, $\boldsymbol{v}_{k,t}$ is the frequency count vector of words generated from this topic at epoch $t$. This a state space model with nonlinear emission, and fortunately there is a large body of literature on how to use Metropolis-Hasting to sample from this distribution (Tanizaki, 2003, Geweke and Tanizaki, 2001). There are two strategy to deal with this posterior: either sample from it as a block, or to sequentially sample each $\phi_{kt}$. Both of these options involve an M-H proposal, however, due to the strong correlation between the successive values of $\phi_{kt}$, we found that sampling this posterior as a block is superior. Let $q(\boldsymbol{\phi_k})$ be the proposal distribution, and let $\boldsymbol{\phi_k^*}$ denote a sample from this proposal. The acceptance ratio is $r = min(1, u)$, where $u$ is as follows (for simplicity assume that the chain starts at $t = t_1$):

$$\frac{H(\phi_{k,t_1}^*) \times \prod_t f(v_{kt}|\phi_{kt}^*) P(\phi_{kt}^*|\phi_{k,t-1}^*)}{H(\phi_{k,t_1}) \times \prod_t f(v_{kt}|\phi_{kt}) P(\phi_{kt}|\phi_{k,t-1})} \times \frac{\prod_t q(\phi_{kt})}{\prod_t q(\phi_{kt}^*)} \quad (14)$$

---
[1] Ahmed and Xing (2008) used a finite dynamic-mixture model, which is equivalent on the limit to RCRP, to compute the same quantity. Our formula here is exact and have the same amount of computation as the approximate formula. We also note that our formula gives better results

With probability $r$ the proposed values are accepted and with probability $1 - r$ the old values are retained. Our proposal is based on a Laplace approximation to the LTR smoother (details of calculating this proposal is given in the appendix). A similar Laplace proposal has been used successively in the context of Bayesian inference of the parameters of a Logistic-Normal distribution (Hoff, 2003), as well as in the context of non-linear state space models in Geweke and Tanizaki (2001) who also noted that this proposal has high acceptance rate (a fact that we also observed).

### 4.1 Practical Considerations

A naive implementation of the Gibbs sampling algorithm in 4 might be slow. We note here that the difference between the sampler we described in 4 and the sampler of a standard CRFP comes in the calculation of the vales of $m'_{kt}$ and for the calculation of (10). The case for $m'$ is simple if we note that it needs to be computed only once before sampling variables in epoch $t$, moreover, because of the form of the exponential kernel used, we can define a recurrence over $m'_{kt}$ as: $m'_{kt} = (m'_{k,t-1} + m_{k,t-1})\exp^{\frac{-1}{\lambda}}$ if $t < \Delta$ and $m'_{kt} = (m'_{k,t-1} + m_{k,t-1})\exp^{\frac{-1}{\lambda}} - \exp^{\frac{-(\Delta+1)}{\lambda}} m_{k,t-(\Delta+1)}$ otherwise.

On the other hand, a naive implementation of (10) costs an $O(K^2\Delta)$ as we need to compute it for $\Delta$ epochs and for each $k$. Here we describe an alternative approach. We divide and multiply (10) with $P(\boldsymbol{k}_{t+\delta}|\boldsymbol{k}_{t+\delta-\Delta:t+\delta-1})$, which is the likelihood of the table assignments at epoch $t + \delta$ given the current configuration with the old value of $k_{tdb} = k^{old}$. This is legitimate since this value is constant across $k$. Now we absorb the value we multiplied in the normalization constant, and focus on the following ratio:

$$C^{t+\delta}(k^{old} \rightarrow k^{new}) = \frac{P(\boldsymbol{k}_{t+\delta}|\boldsymbol{k}_{t+\delta-\Delta:t+\delta-1}^{-tdb\rightarrow k})}{P(\boldsymbol{k}_{t+\delta}|\boldsymbol{k}_{t+\delta-\Delta:t+\delta-1})} \quad (15)$$

where $C^{t+\delta}(k^{old} \rightarrow k^{new})$ is the cost contributed by the assignment of the table at epoch $t + \delta$ for moving an occupancy from table $k^{old}$ to $k^{new}$. In fact (15) is all what we need to compute (10) and thus (9). The idea here is that all the terms not involving $k^{old}$ and $k^{new}$ will cancel from (15), leaving only 2 terms that involve $k^{old}$ and $k^{new}$ to be computed. To see why this is the case, note that $m'^{,-tdb\rightarrow k^{new}}_{s,t+\delta}$ reduces to $m'_{s,t+\delta}$ whenever $s \notin \{k^{new}, k^{old}\}$. Furthermore, we can cache this two dimensional array at each epoch and dynamically update it whenever the sampled value in (9) for $k_{tdb}$ is different from $k^{old}$. In this case, we need to update $C^{t+\delta}(k^{old} \rightarrow .)$ and $C^{t+\delta}(k^{new} \rightarrow .)$. Thus the cost of computing the transition probability reduces from $O(K^2\Delta)$ to at most $O(K\Delta)$. Moreover, especially at later stages of the sampler when tables do not change their topic assignments frequently, the improvement ratio will be more than that.

## 5 Experimental Results

In this section we illustrate iDTM by measuring its ability to recover the death and birth of topics in a simulated dataset and in recovering topic evolution in the NIPS dataset. For all the experiments in this paper, we place a vague gamma prior (1,1) over the hyperparameters $\gamma, \alpha$ and sample them separately for each epoch using the method described in Teh et al. (2006). Unless otherwise stated, we use the following values for the hyperparameters: the variance of the base measure $H$, $\sigma = 10$; the variance of the random walk kernel over the natural parameters of each topic, $\rho = 0.01$; $\Delta = 4$; the number of sample from the base measure $Q = 5$, and finally $\lambda = .5$.

**Initialization** of the Markov chain is quite important. Our setup proceeds as follows. In the first iteration, we run the Gibbs sampler in the filtering mode (i.e. sampling each epoch conditioned on the $\Delta$ preceding epochs only) and used a liberal value for $\alpha = 4$ and $\gamma = 10$ to encourage the generation of large number of topics. An initial large number of topics in desirable since as noted in Teh et al. (2007), initializing HDP-like models with large number of topics results in a better mixing than initializing the sampler with a smaller number of topics. In the subsequent iterations, we ran our standard Gibbs sampler that also samples the values of $\alpha, \gamma$. Finally, we ran all samplers for 2000 iterations and took 10 samples 200 iterations apart and then used the sample with the highest Likelihood for evaluation and visualization

### 5.1 Simulation Results

We generated a simple time-evolving document collection over $T = 20$ epochs. We set the vocabulary size to 16, and hand-crafted the birth-death of 8 topics, as well as their words' distributions as shown in Figure 2.

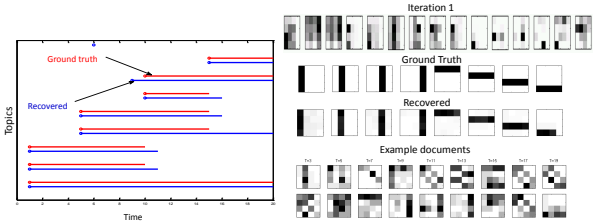

Figure 2: Illustrating simulation results. **Left**: topic's death-birth over time (topics numbered from bottom-top). Ground truth is shown in red and recovered in blue. **Right**: from top to bottom, topics' distribution after iteration 1, a posterior sample, ground truth (numbered from left to right), and finally a set of documents at different time epochs from the training data.

Each topic puts its mass on 4 words. At each epoch we add a 5% random noise to each topic's word distribution. We then ran the RCRFP with $\alpha = 1.5$ to generate 100 documents at each epoch each of which having 50 words. $\gamma$ was set to zero in this generation since the topics layout were fixed by hand. Moreover, Once a topic is alive, say at epoch $t$, its prior popularity $m'$ is set to the average prior popularity at epoch $t$. Our goal was to assess the efficacy of iDTM in recovering abrupt death and birth of topics. Finally given the generated data, we ran the sampler described in Section 4 to recover the topics and their durations. As depicted in Figure 2, iDTM was able to recover the correct distribution for each topic as well as its correct lifespan.

### 5.2 Timeline of the NIPS Conference

We used iDTM to analyze the proceedings of the NIPS conference from the years 1987-1999. We removed words that appear more than 5000 times or less than 100 times which results in a vocabulary size of 3379 words. The collection contains 1740 documents, where each document contains on average 950 words. Documents were divided into 13 epochs based on the publication year. We ran iDTM to recover the structure of topic evolution in this corpus.

Figure 3.a shows the initial state of the sampler and the MAP posterior sample. Each horizontal line gives the duration of a topic where the x-axis denotes time. Figure 3.c shows the number of topics in the collection over time. We also draw the symmetrized KL-divergence between the unigram distribution of words at epoch $t$ and $t-1$. It can be noticed that whenever there is a sharp change in the KL value, the model responds by changing the number of topics. However, when the KL value is stable (but not zero), the model responds by changing the word distributions of the topics and/or the topics' trends. This is in contrast to DTM which can only change the last two quanti-

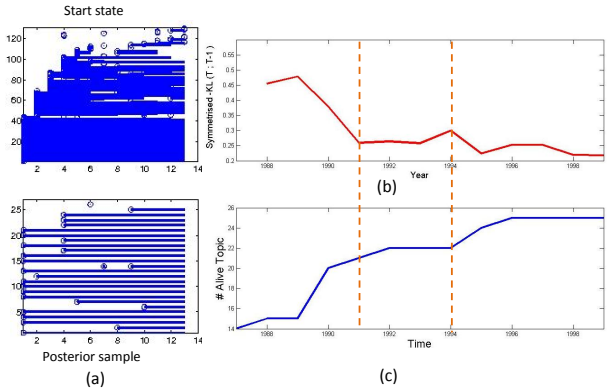

Figure 3: **(a)** The sampler initial state and the MAP posterior sample. Each line represents the lifespan of a topic. **(b)** Symmetrized-KL divergence between the unigram distribution of words at epoch $t, t-1$. **(c)** The number of alive topics over years in the NIPS collections.

ties. We would like to add that the trend of always-increasing number of topics is not an artifact of the model, but rather a property of the NIPS conference in this lifespan: none of the topics that we observed die completely during this time period. Moreover, as we illustrated in the simulation study, the model can detect an abrupt death of topics.

In Figure 4, we show a *timeline* of the conference pointing out the birth of some of the topics. We also give how their trends change over time and show a few examples of how the top words in each topic change over time. In Figure 5, we show a *timeline* of the Kernel topic illustrating, in some years, the top 2 (3 in case of a tie) papers with the highest weights for this topic. Indeed the three papers in 1996 are the papers that started this topic in the NIPS community. We would like to warn here that the papers having the highest weights of a topic need not be the most influential papers about this topic. Perhaps this is true in the year in which the topic was born, but for subsequent years, these papers give an overview of how this topic is being addressed along the years, and it can provide a concise input for summarization systems utilizing topic models as in (Haghighi and Vanderwende, 2009). Finally it is worth mentioning that iDTM differs from DTM in the way they model topic trends: DTM assumes a smooth evolution of trends, whereas iDTM assumes a non-parametric model and as such can spawn a topic with a large initial trend as in the Kernel topic in 1996.

#### 5.2.1 Quantitative Evaluation

In Addition to the above qualitative evaluation, we compared iDTM against DTM (Blei and Lafferty, 2006) and against HDP (Teh et al., 2006). To compare with DTM, we followed the model in (Blei and

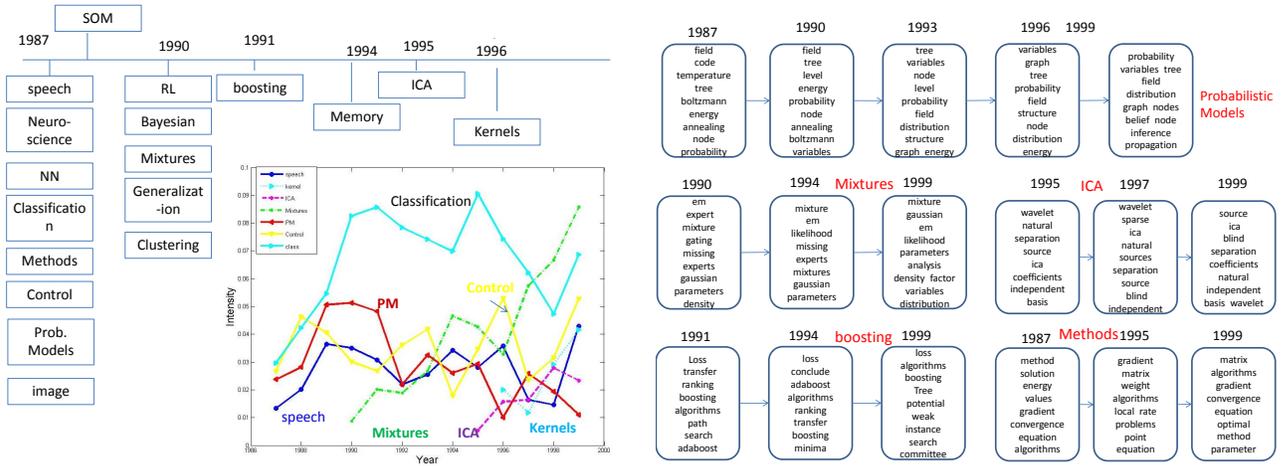

Figure 4: Timeline for the NIPS conference. **Left**: birth of a number of topics and their trends over time. **Right**: top words in some topics over time.

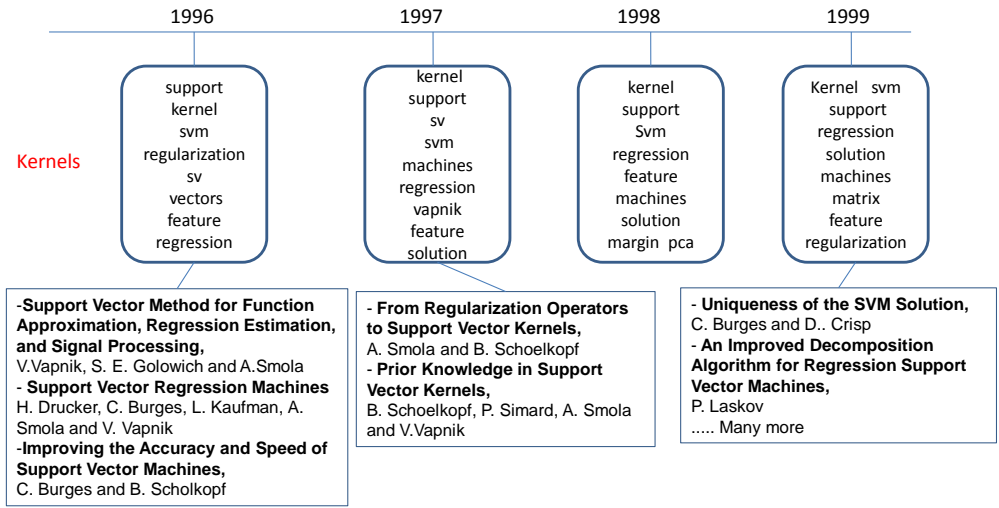

Figure 5: Timeline for the Kernel topic. The figure shows the top words in the topic in each year and the top 2 (3 in case of a tie) papers with the highest weights of this topic in some years

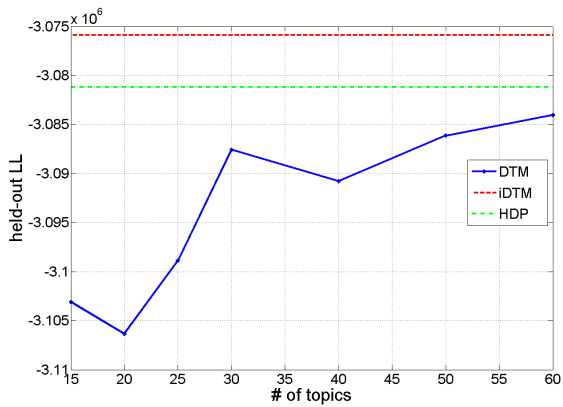

Figure 6: Held-out LL comparison between DTM, iDTM and HDP.

Lafferty, 2006), and used a diagonal covariance for the logistic-normal distribution over the topic-mixing vector at each epoch. We linked the means of the logistic-normal distributions at each epoch via a random walk model, and we evolved the distribution of each topic as we did in iDTM. To make a fair comparison, in fitting the variational distribution over the topic's word distribution $\phi$ in DTM, we used a Laplace variational approximation similar to the one we used in fitting the proposal distribution for iDTM. Moreover, we used importance sampling with the variational distribution as a proposal for calculating the test LL for DTM.

We divided the data into 75% for training and 25% for testing, where the training and test documents were selected uniformly across epochs. As shown in Figure 6, iDTM gives better predictive LL over DTM and HDP.

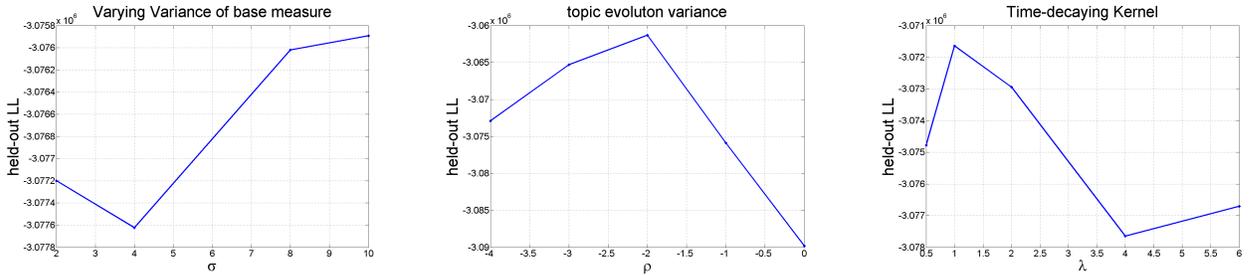

Figure 7: Sensitivity of iDTM to hyperparameters. Every panel vary one hyperparameter while keeping all other hyperparameters fixed to their default values. **Left**: Varying base measure variance $\sigma$. **Middle**: Variance of the random walk model over topic parameters $\rho$ ($\rho$ is drawn in log-scale). **Right**: Parameter of time-decaying kernel $\lambda$ ($\Delta$ is fixed at 13 in this specific experiments)

### 5.2.2 Hyperparameter Sensitivity

To assess the sensitivity of iDTM to hyperparameters' settings, we conducted a sensitivity analysis in which we hold all hyperparameters fixed at their default values, and vary one of them. As noted earlier, the hyperparameters are: the variance of the base measure $\sigma$; the variance of the random walk kernel over the natural parameters of each topic, $\rho$; and the parameter of the time-decaying kernel, $\lambda$. We should note here that the order of the process $\Delta$ can be safely set to $T$, however, to reduce computation, we can set $\Delta$ to cover the support of the time-decaying kernel, i.e, we can choose $\Delta$ such that $\exp^{\frac{-\Delta}{\lambda}}$ is smaller than a threshold, say .001. The results are shown in Figure 7.

When $\rho$ is set to 1, the performance deteriorates and the topics become incoherent over time. We noticed that in this setting the model recovers only 5 to 7 topics. When $\rho$ is set to .0001, the word distribution of each topic becomes almost fixed over time. In between, the model peaks at $\rho = .01$. It should be clear from the figure that an underestimate of the optimal value of $\rho$ is less detrimental than an overestimate, thus we recommend setting $\rho$ in the range $[.001, .1]$. It should be noted that we could add a prior over $\rho$ and sample it every iteration as well; we leave this for future work.

While varying $\lambda$, we fixed $\Delta = T$ to avoid biasing the result. A large value of $\lambda$ degenerates the process toward HDP, and we noticed that when $\lambda = 6$, some topics,like ICA, weren't born and where modeled as a continuation of other related topics. $\lambda$ depends on the application and the nature of the data. In the future, we plan to place a discrete prior over $\lambda$ and sample it as well. Finally, the best setting for the variance of the base measure $\sigma$ is from $[5, 10]$, which results in topics with reasonably sparse word distributions.

## 6 Conclusions and Future work

In this paper we addressed the problem of modeling time-varying document collections. We presented a topic model, iDTM, that can adapt the number of topics, the word distributions of topics, and the topics' trend over time. To the best of our knowledge, this is the first model of its kind. We used the model to analyze the NIPS conference proceedings and drew several timelines for the conference: a timeline of topic birth and evolution as well as a timeline for each topic that shows its trend over time and the papers with the highest weight of this topic in its mixing vector. This information provides a bird's eye view of the collection, and can be used as input to a summarization system for each topic. In the future, we plan to extend our Gibbs sampler to sample all the hyperparameters of the model. We also plan to extend our model to evolve an HDP at various levels, for instance, lower levels might correspond to conferences, and the highest level to time. This framework will enable us to understand topic evolution within and across different conferences or disciplines.

## 7 Acknowledgment

This work is supported in part by grants NSF IIS-0713379, NSF DBI-0546594 career award to EPX, ONR N000140910758, DARPA NBCH1080007. EPX is supported by an Alfred P. Sloan Research Fellowship. We thank the anonymous reviewers for their helpful comments

## References

D. Blei, A. Ng, and M. Jordan. Latent dirichlet allocation. *Journal of Machine Learning Research*, 3:993–1022, 2003.

D. Blei and J. Lafferty. A correlated topic model of science. *Annals of Applied Statistics*, 1(1):17–35, 2007.

W. Li and A. McCallum. Pachinko allocation: Dag-structured mixture models of topic correlations. *ICML*, 2006.

D. Blei and J. Lafferty. Dynamic topic models. In *ICML*, 2006.

X. Wang and A. McCallum. Topics over time: A non-markov continuous-time model of topical trends. In *KDD*, 2006.

A. Ahmed and E.P. Xing. Dynamic non-parametric mixture models and the recurrent chinese restaurant process with application to evolutionary clustering. In *SDM*, 2008.

F. Caron, M. Davy, and A. Doucet. Generalized polya urn for time-varying dirichlet processes. In *UAI*, 2007.

N. Srebro and S. Roweis. Time-varying topic models using dependent dirichlet processes. In *Technical report, Department of Computer Science, University of Toronto*, 2005.

T. S. Ferguson. A bayesian analysis of some nonparametric problems. *The Annals of Statistics*, 1(2):209–230, 1973.

D. Blackwell and J. MacQueen. Ferguson distributions via polya urn schemes. *The Annals of Statistics*, 1(2):353–355, 1973.

C. E. Antoniak. Mixtures of dirichlet processes with applications to bayesian nonparametric problems. *The Annals of Statistics*, 2(6):1152–1174, 1974.

X. Zhu Z. Ghahramani and J. Lafferty. Time-sensitive dirichlet process mixture models. In *Technical Report CMU-CALD-05-104*, 2005.

J.E. Griffn and M.F.J. Steel. Order-based dependent dirichlet processes. *Journal of the American Statistical Association*, 101(473):1566–1581, 2006.

Y. Teh, M. Jordan, M. Beal, and D. Blei. Hierarchical dirichlet processes. *Journal of the American Statistical Association*, 101(576):1566–1581, 2006.

R. M. Neal. Markov chain sampling methods for dirichlet process mixture models. *Technical Report 9815, University of Toronto, Department of Statistics and Department of Computer Science*, 1998.

H. Tanizaki. Nonlinear and non-gaussian state-space modeling with monte carlo techniques: A survey and comparative study. In *Handbook of Statistics Vol.21, Stochastic Processes: Modeling and Simulation, Chap.22, pp.871-929 (C.R. Rao and D.N. Shanbhag, Eds.)*, 2003.

J. Geweke and H. Tanizaki. Bayesian estimation of state-space model using the metropolis-hastings algorithm within gibbs sampling. *Computational Statistics and Data Analysis*, 37(2):151–170, 2001.

P. Hoff. Nonparametric modeling of hierarchically exchangeable data. In *UW Statistics Department Technical Report no. 421*, 2003.

Y.W. Teh, K. Kurihara, and M. Welling. Collapsed variational inference for hdp. In *NIPS*, 2007.

A. Haghighi and L. Vanderwende. Exploring content models for multi-document summarization. In *HLT-NAACL*, 2009.

T. Minka From Hidden Markov Models to Linear Dynamical Systems. TR 53, 1998.

V. Rao and Y.W. Teh Spatial Normalized Gamma Processes. NIPS 2009.

## Appendix A: Fitting the proposal distribution in (14)

Recall that our goal is to sample $P(\boldsymbol{\phi_k}|\boldsymbol{v_k})$, where $\boldsymbol{v_k} = \{\boldsymbol{v_{k,t}}\}$, $\boldsymbol{v_{k,t}}$ is the frequency count vector of words generated from this topic at epoch $t$. Without loss of generality, and for notational simplicity, we will drop the topic index $k$, and assume that the topic's lifespan is form 1 to $T$. Thus we would like to compute $P(\phi_1, \ldots, \phi_T | v_1, \ldots, v_T)$. This is a linear state-space model with non-linear emission, thus the RTS smoother (Minka, 1998) will not result in a closed form solution. Therefore, we seek a Laplace-approximation to this posterior and call this approximation, $q(\phi) = \prod_t q(\phi_t)$. To compute $q$ we note that the RTS smoother defines two recurrences: the forward recurrence, and the backward recurrence. Following Minka (1998), lets assume that the forward value at epoch $t-1$ is given by $\hat{\alpha}(\phi_{t-1}) \sim \mathcal{N}(\mathbf{u}_{t-1}, \boldsymbol{\Upsilon}_{t-1})$ where $\boldsymbol{\Upsilon}$ has a diagonal covariance. The forward equation becomes:

$$\begin{aligned}
\hat{\alpha}(\phi_t) &= \mathcal{LN}(v_t|\phi_t) \times \\
&\quad \int_{\phi_t} \mathcal{N}(\phi_t|\phi_{t-1}, \rho I) \mathcal{N}(\phi_{t-1}|\mathbf{u}_{t-1}, \boldsymbol{\Upsilon}_{t-1}) \\
&= \mathcal{LN}(v_t|\phi_t) \mathcal{N}(\phi_t|\phi_{t-1}, \boldsymbol{\Upsilon}_{t-1} + \rho I) \quad (16)
\end{aligned}$$

Equation (16) does not result in the desired Gaussian form because of the non-conjugacy between the $\mathcal{LN}$ and the normal distributions. Therefore, we seek a Laplace approximation to (16) which puts it back into the desired Gaussian form to continue the recurrence. In this case $\mathbf{u}_t$ is the mode of (16) and $\boldsymbol{\Upsilon}_t$ is the negative inverse Hessian of (16) evaluated at the mode. We use a Diagonal-approximation of the Hessian though because of the high-dimensionality of $\phi$.

As detailed in Minka (1998), the backward recurrence can be defined using the $\hat{\alpha}$'s instead of the data, and can be computed exactly if the dynamic model is linear, which is the case in our model. This backward recurrence computes $q(\phi_t|v_1, \ldots, v_T)$ that we desire.